\documentclass[aps,reprint,prd,showpacs,amsmath,amssymb,nofootinbib]{revtex4-1}
\usepackage{graphicx}
\usepackage{color}
\usepackage{times}
\usepackage{inputenc}
\usepackage{bm}
\usepackage{multirow}
\usepackage{float}
\usepackage{url}
\usepackage{natbib}
\usepackage{tensor}
\usepackage{xcolor}
\usepackage[colorlinks=true,citecolor=blue,urlcolor=blue,linkcolor=red]{hyperref}
\usepackage[normalem]{ulem}

\def\msun{\,M_{\odot}}
\def\fm3{\;\text{fm}^{-3}}
\def\mev{\;\text{MeV}}

\begin{document}
\title{Nonstrange and strange quark matter at finite temperature within modified NJL model and protoquark stars}

\author{Wen-Li Yuan$^{1}$}
\email{wlyuan@pku.edu.cn} 
\author{Nobutoshi Yasutake $^{2}$}
\email{nobutoshi.yasutake@it-chiba.ac.jp}
\author{Renxin Xu$^{1}$}
\email{r.x.xu@pku.edu.cn}

\affiliation{$^1$School of Physics and State Key Laboratory of Nuclear Physics and Technology, Peking University, Beijing 100871, China;\\ 
$^2$Department of Physics, Chiba Institute of Technology (CIT), 2-1-1 Shibazono, Narashino, Chiba 275-0023, Japan
}

\begin{abstract}

We extend the modified Nambu–Jona-Lasinio (NJL) model--incorporating exchange interactions via the Fierz transformation--to finite temperatures in both two- and three-flavor scenarios, and investigate the properties of protoquark stars in $\beta$-equilibrium. Our results show that increasing the strength of exchange interactions, characterized by the parameter $\alpha$, changes the chiral phase transition from first-order to crossover. We examine the effects of finite temperature, lepton fraction, and exchange interactions on the equation of state (EOS). We find that, in the crossover regime, the EOS is significantly stiffer than in the first-order case due to the substantial contribution of repulsive interactions in the exchange channels, while it remains relatively insensitive to variations in temperature and lepton fraction. Imposing the thermodynamic consistency,  which requires the minimum of $f / \rho_B$ occurs at zero pressure, further constrains the minimum value of vacuum pressure.

\end{abstract}

\maketitle

\section{Introduction}
Neutron stars are the remnants of supernova explosions. They typically have masses in the range of $1-2\msun$, radii on the order of $\sim 10$~km, and are born with temperatures as high as $10^{11}$~K. Within a few days, they cool to approximately $10^{10}$~K by emitting neutrinos. In stellar modeling, the internal structure of a neutron star is determined by the assumed equation of state (EOS), which remains uncertain due to the incomplete understanding of the strong interaction under extreme conditions. At present, extensive studies have been conducted on proto-neutron stars with or without quark cores at finite temperature~\cite{1997PhR...280....1P,1998PhRvD..58a3009R,2000PhLB..486..239S,2001PhRvL..86.5223P,2001PhLB..509...10S,2006PhRvD..74l3001N,2008PhRvD..77h5022B,2009PhRvD..79d3012Y,2011PhLB..704..343S,2012PhRvD..86d5006C,2015EPJA...51..155M,2019ApJ...875...12R}.

According to the Bodmer-Witten hypothesis~\cite{1971PhRvD...4.1601B,1984PhRvD..30..272W}, compact stars can be composed entirely of deconfined quarks, forming self-bound quark stars. The composition of these self-bound stars hinges on a fundamental question: what is the true ground state of baryonic matter at zero pressure and temperature? After decades of theoretical speculation, both strange quark stars, composed of strange quark matter~\cite{1989PhLB..229..112C,1998PhLB..438..123D,1993PhRvD..48.1409C,1999PhLB..457..261B,1999PhRvC..61a5201P,2005PhRvC..72a5204W,2010MNRAS.402.2715L,2018PhRvD..97h3015Z,2018PhRvD..98h3013L,2019PhRvD..99j3017X,2000PhRvC..62b5801P,2021EPJC...81..612B,2021ChPhC..45e5104X,2022PhRvD.105l3004Y,2024PhRvD.110j3012Z,2024FrASS..1109463Z}, and up-down quark stars, containing only up and down quark matter~\cite{2018PhRvL.120v2001H,2019PhRvD.100d3018Z,2020PhRvD.101d3003Z,2021PhRvD.103f3018Z,2023PhRvD.107k4015R,2024Ap&SS.369...29S,2025arXiv250700776Y}, have emerged as viable alternative models to neutron stars. Besides, self-bound stars composed of strangeon matter have also been proposed as candidates~\cite{2003ApJ...596L..59X,2009MNRAS.398L..31L,2023PhRvD.108f3002Z,2025PhRvD.111f3033Y}. Potential candidates for quark stars include compact objects with gravitational masses $\lesssim 1 \msun$, such as PSR J1645-0317 (PSR B1642-03)~\cite{1993ApJS...88..529T}, 
PSR J1830-1059 (PSR B1828-11)~\cite{1993ApJS...88..529T}, 4U 1728-34~\cite{1999ApJ...527L..51L}, and the X-ray bursters GRO J1744-28~\cite{1998Sci...280..407C} and SAX J1808.4-3658~\cite{1999PhRvL..83.3776L}. A recent analysis of the supernova remnant HESS J1731-347 suggests that its central compact object has a remarkably low mass of $M = 0.77^{+0.20}_{-0.17}\msun$ and a radius of $R = 10.4^{+0.86}_{-0.78}$ ~km~\cite{2022NatAs...6.1444D}, reigniting interest in the possible existence of quark stars. While observations of isolated compact stars provide valuable constraints on the EOS of cold dense matter, many dynamical astrophysical phenomena--such as core-collapse supernovae, the formation and cooling of proto-compact stars, and compact star mergers--depend on the EOS at finite temperatures. To account for thermal effects, an ideal-fluid thermal component is commonly introduced into an arbitrary cold EOS~\cite{1993A&A...268..360J}. In this prescription, the thermal contribution is parameterized by a simple adiabatic index as $P_{th}=\epsilon_{th} (\Gamma_{th}-1)$, where $P_{th}$ and $\epsilon_{th}$ are the thermal pressure and energy density, and $\Gamma_{th}$ is the adiabatic index, which is assumed to be constant. Although this treatment offers computational simplicity, it neglects the effects of degeneracy on the thermal pressure and lacks explicit dependence on temperature and entropy. While several nuclear-matter EOSs have been developed to incorporate finite-temperature effects self-consistently~\cite{1991NuPhA.535..331L,1998NuPhA.637..435S,2010NuPhA.837..210H,2025arXiv250617569Z}, only a limited number of them exist, and even fewer account for phase transitions or exotic degrees of freedom. Therefore, further investigation of thermal effects in quark matter and the evolution of protoquark stars within effective field theory frameworks remains both essential and intellectually compelling~\cite{2006JPhG...32.1081M,2013EPJC...73.2569D,2014JPhG...41a5203D,2017EPJC...77..512C,2019PhRvD.100j3012C}.

Due to the nonperturbative challenges at finite baryon densities, effective models that incorporate key features of QCD offer a practical alternative. 
For example, Ref.~\cite{2021PhyS...96f5302L} employed the modified MIT bag model to study the quark matter at finite temperature and hot stars. To effectively capture the dynamics of spontaneous chiral symmetry breaking in QCD, the Nambu-Jona-Lasinio (NJL) model~\cite{1992RvMP...64..649K,2005PhR...407..205B} serves as a particularly suitable choice, which successfully reproduces the spectrum of low-lying mesons~\cite{2005PhR...407..205B,1992RvMP...64..649K}, and has been widely applied and extended to describe quark matter in compact star physics (see, e.g., Refs.~\cite{2003PhRvD..67f5015H,2005PhRvD..72f5020B,2005PhRvD..72f5020B,2019ChPhC..43h4102W,2021PhRvC.104f5201M,2024PhRvD.110a4022X,2025PhRvD.111a4006G,2024arXiv241104064G} as incomplete lists). Within this framework, Ref.~\cite{2017EPJC...77..512C} investigates the role of isovector interactions and the behavior of the symmetry free energy of strange quark matter at finite temperature. 
In Ref.~\cite{2006JPhG...32.1081M}, the authors compare the predictions derived from the standard NJL model containing scalar interactions with those of the MIT bag model for strange quark matter EOS at finite temperature, as well as the properties of strange quark stars.

To improve the description of strongly interacting matter at high baryon chemical potentials, a modified version of the NJL model has been proposed in Ref.~\cite{2019ChPhC..43h4102W}. This extended NJL model incorporates both the original NJL Lagrangian and its Fierz-transformed counterpart, with their respective contributions weighted by the parameters $1-\alpha$ and $\alpha$. The parameter $\alpha$ is treated as a free parameter, allowed to vary between $0$ to $1$, enabling the model to accommodate constraints at finite baryon density.  In this sense, the present formulation generalizes the NJL-type Lagrangian, of which the specific choice $\alpha=0.5$ introduced in Ref.~\cite{1992RvMP...64..649K} represents a special case. This modified NJL model has been successfully applied in various studies, including the study of the chiral chemical potential $\mu_5$ and its physical effects in two-flavor quark matter system~\cite{2019PhRvD.100i4012Y,2021ChPhC..45h4110W}, investigations of the critical endpoint in the QCD phase diagram within the three-flavor version of the model and its comparison with the two-flavor case~\cite{2020ChPhC..44g4104Y}, as well as the the absolute stability and phase structure of two-flavor color superconductivity~\cite{2020PhRvD.102e4028S,2023PhRvD.108d3008Y,2024ApJ...966....3Y,2025arXiv250706676Y}. Compact stars, serving as natural laboratories to explore the nature of dense matter, have strongly motivated studies of nonstrange and strange quark stars~\cite{2020MPLA...3550321W,2022PhRvD.105l3004Y,2024Ap&SS.369...29S}, as well as neutron stars with quark-matter cores~\cite{2024arXiv240805687Y,2025PhRvD.112b3019Y}, within the framework of the modified NJL model with exchange interactions parameterized by $\alpha$ to describe the cold quark matter phase. To our knowledge, existing studies within the NJL-like model have primarily focused on cold quark matter~\cite{2020MPLA...3550321W,2022PhRvD.105l3004Y,2024Ap&SS.369...29S} or thermal three-flavor quark matter systems~\cite{2006JPhG...32.1081M,2017EPJC...77..512C}. However, the case of nonstrange quark matter and its systematic comparison with strange quark matter at finite temperature--particularly regarding the properties of nonstrange and strange protoquark stars--has not yet been explored in detail. Extending these earlier efforts, in this work, we employ this modified NJL-type model to describe two- and three-flavor quark matter with and without trapped neutrinos at finite temperature. We then systematically investigate the impact of exchange interactions on the characteristics of chiral phase transition and explore their implications for both strange and nonstrange protoquark stars.

This paper is organized as follows. In Sec.~\ref{Sec: formulalism}, we will present the modified NJL model employed in this work to describe the nonstrange and strange quark matter at finite temperature, including the Fierz transformed interactions. Sections~\ref{Sec3: quark matter at finite T},~\ref{Sec: QM finite tem} and Sec.~\ref{Sec: star} discuss the results on quark matter EOS with and without trapped neutrinos at finite temperature and protoquark stars. Our results are summarized in Sec.~\ref{Sec: summary}.

\section{The modified NJL models}\label{Sec: formulalism}
In this section, we introduce the modified NJL model incorporating exchange channels via Fierz transformations, which account for the effects of rearranging fermion field operators.

\subsubsection{two-flavor modified NJL model}\label{Sec: 2f NJL model}
The Lagrangian of the two-flavor NJL model reads:
\begin{equation}
\begin{aligned}
\mathcal{L}_{\mathrm{NJL}}^{~2f} =&\mathcal{L}_{0}+G\left[(\bar{\psi} \psi)^{2}+\left(\bar{\psi} i \gamma^{5} \tau \psi\right)^{2}\right] \ , \\
\mathcal{L}_{0}=&\bar{\psi}\left(i \gamma^{\mu}\partial_{\mu} -m +\mu \gamma^{0} \right)\psi
\ . \label{eq1}
\end{aligned}
\end{equation}
Here, $\mathcal{L}_{0}$ denotes the relativistic free field, which describes the propagation of non-interacting fermions. $G$ is the four-fermion interaction coupling constant. $\psi$ is the quark field operator with color, flavor, and Dirac indices. $\mu$ is the flavor-dependent quark chemical potential. The quark mass matrix $m = \mathrm{diag}(m_u, m_d)$ is diagonal in flavor space and includes the small current quark masses, thereby introducing a slight explicit breaking of chiral symmetry.

For further considering the effect of a rearrangement of fermion field operators, we apply the Fierz transformation to the interaction terms in the NJL models~\cite{2019ChPhC..43h4102W}. As a purely technical device to examine the exchange channels influence that occur in quartic products at the same space-time point~\cite{1992RvMP...64..649K,2005PhR...407..205B}, the Fierz identity of the four-fermion interactions in the two-flavor NJL model is
\begin{equation}
\begin{aligned}
\mathcal{F}(\mathcal{L}_{\sigma}^{4})=& \frac{G}{24}\left[2(\bar{\psi} \psi)^{2}+2\left(\bar{\psi} i \gamma^{5} \tau \psi\right)^{2}-2(\bar{\psi} \tau \psi)^{2}\right.\\
&-2\left(\bar{\psi} i \gamma^{5} \psi\right)^{2}-4\left(\bar{\psi} \gamma^{\mu} \psi\right)^{2}-4\left(\bar{\psi} i \gamma^{\mu} \gamma^{5} \psi\right)^{2} \\
&\left.+\left(\bar{\psi} \sigma^{\mu \nu} \psi\right)^{2}-\left(\bar{\psi} \sigma^{\mu \nu} \tau \psi\right)^{2}\right] \ , \label{Fierz_2f_inter}
\end{aligned}
\end{equation}
with only considering the contribution of color singlet terms for simplicity. One can see that, through the Fierz transformation, all exchange interaction channels of the original Lagrangian are released. In Eq.~(\ref{Fierz_2f_inter}), the Fierz transformed Lagrangian contains not only the scalar and pseudoscalar interactions, but also vector and axial-vector interaction channels.

Due to the mathematical equality between the original interactions and Fierz-transformed interactions, we can combine them using a weighting factor $\alpha$. The factor $\alpha$ reflects the competition between the original interaction channels and the exchange interaction channels.
Then the effective Lagrangian becomes
\begin{equation}
\mathcal{L}_{\rm eff}^{~2f}=\bar{\psi}(i \gamma^{\mu}\partial_{\mu} -m+\mu\gamma^{0}) \psi +(1-\alpha)\mathcal{L}_{\mathrm{int}}^{~2f}+\alpha \mathcal{F}(\mathcal{L}_{\mathrm{int}}^{~2f}) \ .\label{eq7}
\end{equation}
When $\alpha = 0$, the effective Lagrangian reduces to the standard NJL model. Unlike the conventional approach of artificially adding the vector interaction to the standard NJL Lagrangian for systems at finite chemical potential, the modified Lagrangian in Eq.~(\ref{eq7}) introduces vector interactions self-consistently~\cite{2019ChPhC..43h4102W}. 

Under the mean-field approximation, the scalar and vector interactions in the Fierz-transformed channels in Eq.~(\ref{Fierz_2f_inter}) also contribute at finite densities, thereby modifying the constituent quark mass $M$ and the effective quark chemical $\tilde{\mu}$. The corresponding mass gap equation and the effective chemical potential for each quark are given as follows:
\begin{equation}
\begin{aligned}
M=&m -2\left[(1-\alpha) +\frac{\alpha }{12} \right]G \sum_{f=u, d}\sigma_f \\
 =&m-2 G' \sum_{f=u, d}\sigma_f \ ,\\
 \tilde{\mu} =&\mu-\frac{\alpha}{3} G\sum_{f=u, d}\rho_f\ ,
\end{aligned} \label{eq: 2f gap Eq}
\end{equation}
where we define $G'=(12-11\alpha)G/12$. The quark condensate $\langle\bar{\psi} \psi\rangle$ and quark number density $\left\langle\psi^{+} \psi\right\rangle$ are denoted as $\sigma$ and $\rho$, respectively, which are the average values of operators, $\bar{\psi} \psi$ and $\psi^{+} \psi$, in the ground state. 

In finite-temperature field theory~\cite{2011ftft.book.....K}, the linearization
of $\mathcal{L}_{\rm eff}$ in the vicinity of the expectation values and the application of Matsubara formalism yield the thermodynamical potential per volume of quark matter expressed as:
\begin{equation}
\begin{aligned}
&\Omega_M(T, \tilde{\mu})=\\
&-  6 \sum_{f=u, d} \int \frac{\mathrm{d}^3 p}{(2 \pi)^3}  \left[E_f +T \ln \left(1+e^{-\left(E_f+\tilde{\mu}_f\right) / T}\right)\right. \\
 &\left.+T \ln \left(1+e^{-\left(E_f-\tilde{\mu}_f\right) / T}\right)\right] 
 +2 G'\left(\sigma_u{ }^2+\sigma_d{ }^2\right) \\
 &-\frac{\alpha G}{6}\left(\rho_u{ }^2+\rho_d{ }^2\right)
+\Omega_0 \ .  \label{eq: 2f grand thermo potential}
\end{aligned}
\end{equation} 
From the grand canonical potential density in Eq.~(\ref{eq: 2f grand thermo potential}), all thermodynamic quantities of interest can be calculated using standard thermodynamic relations. 
The pressure is 
\begin{equation}
\begin{aligned}
&P= -\Omega \\
& =6 \sum_{f=u, d} \int \frac{\mathrm{d}^3 p}{(2 \pi)^3}  \left[E_f +T \ln \left(1+e^{-\left(E_f+\tilde{\mu}_f\right) / T}\right)\right. \\
 &\left.+T \ln \left(1+e^{-\left(E_f-\tilde{\mu}_f\right) / T}\right)\right] 
 -2 G'\left(\sigma_u{ }^2+\sigma_d{ }^2\right) \\
 &+\frac{\alpha G}{6}\left(\rho_u{ }^2+\rho_d{ }^2\right) 
- \Omega_0 \ ,\label{eq:pressure}
\end{aligned}
\end{equation}
and the energy density is
\begin{equation}
\begin{aligned}
& \epsilon =-P+T S+\sum_f \mu_f n_f \\
&= 6 \sum_{f=u, d} \int \frac{\mathrm{d}^3 p}{(2 \pi)^3} 
 \left[E_f\left(\bar{n}_f+n_f-1\right) \right]\\
&+2 G'\left(\sigma_u{ }^2+\sigma_d{ }^2\right)+\frac{\alpha G}{6}\left(\rho_u{ }^2+\rho_d{ }^2\right)+\Omega_0\ , \label{eq:energy density}
\end{aligned}
\end{equation}
where $n_f$ and $\bar{n}_f$ are Fermi occupation numbers of quarks and antiquarks, respectively,
\begin{equation}
\begin{aligned}
& n_f\left(T, \tilde{\mu}\right)=\frac{1}{1+\exp \left(\frac{E_f-\tilde{\mu}}{T}\right)}\ ,\\
& \bar{n}_f\left(T, \tilde{\mu}\right)=\frac{1}{1+\exp \left(\frac{E_f+\tilde{\mu}}{T}\right)}\ ,
\end{aligned}
\end{equation}
with $E_f=\sqrt{p^2+M_f^2}$.

\subsubsection{three-flavor modified NJL model}\label{Sec: 3f NJL model}
The Lagrangian density of the three-flavor NJL model is given by  
\begin{equation}
\begin{aligned} 
&\mathcal{L}_{\mathrm{NJL}}^{~3f} =\mathcal{L}_{0}+\mathcal{L}_{\mathrm{int}}^{~3f} \ , \\
&\mathcal{L}_{\mathrm{int}}^{~3f} =  \mathcal{L}_{\sigma}^{4}+ \mathcal{L}_{\sigma}^{6}  \ ,\\
&\mathcal{L}_{\sigma}^{4} =\sum_{i=0}^{8} G\left[\left(\bar{\psi} \lambda_{i} \psi\right)^{2}+\left(\bar{\psi} i\gamma^{5}\lambda_{i} \psi\right)^{2}\right] \ ,\\ 
&\mathcal{L}_{\sigma}^{6} = -K\left(\operatorname{det}\left[\bar{\psi}\left(1+\gamma^{5}\right) \psi\right]+\operatorname{det}\left[\bar{\psi}\left(1-\gamma^{5}\right) \psi\right]\right) \ ,
\label{eqNJL}
\end{aligned}
\end{equation}  
where $\mathcal{L}_{\sigma}^{4}$ and $\mathcal{L}_{\sigma}^{6}$ represent the four-fermion and six-fermion interaction terms, respectively. $G$ and $K$ denote the coupling constants for the four-fermion and six-fermion interactions. The matrices $\lambda_{i}\; (i=1 \rightarrow 8)$ correspond to the Gell-Mann matrices in flavor space. $\lambda_{0} = \sqrt{2/3}\; I_{0}$, with $I_{0}$ being the identity matrix.  
Applying the Fierz transformation to the four-fermion scalar and pseudoscalar interaction terms, considering only the contributions from color-singlet channels, yields  
\begin{equation}
\mathcal{F}(\mathcal{L}_{\sigma}^{4})=-\frac{G}{2}\left[\left(\bar{\psi} \gamma_{\mu} \lambda_{i}^{0} \psi\right)^{2}-\left(\bar{\psi} \gamma_{\mu} \gamma_{5} \lambda_{i}^{0} \psi\right)^{2}\right] \ .\label{eqFierzscalar}
\end{equation}  
Since the Fierz transformation of the six-fermion interaction is defined to preserve invariance under all possible permutations of the quark spinors $\psi$ appearing in the interaction~\cite{1992RvMP...64..649K}, the six-fermion term remains unchanged:  
\begin{equation}
\mathcal{F}(\mathcal{L}_{\sigma}^{6})=\mathcal{L}_{\sigma}^{6} \ .\label{eqFierzSix}
\end{equation}  
Thus, the effective Lagrangian takes the form  
\begin{equation}
\mathcal{L}_{\rm eff}^{~3f}=\bar{\psi}(i \gamma^{\mu}\partial_{\mu} -m+\mu\gamma^{0}) \psi +(1-\alpha)\mathcal{L}_{\mathrm{int}}^{~3f}+\alpha \mathcal{F}(\mathcal{L}_{\mathrm{int}}^{~3f}) \ .\label{eqNJLeff}
\end{equation}  

In the mean-field approximation, the mass gap equations and the effective chemical potential $\mu_{f}^{*}$ for a given flavor $(f=i,j,k)$ are expressed as
\begin{equation}
\begin{aligned}
M_{i}&= m_{i}-4 (1-\alpha) G \sigma_{i}+2 K \sigma_{j} \sigma_{k}\\
&= m_{i}-4G'\sigma_{i}+2 K \sigma_{j} \sigma_{k} , \\
\tilde{\mu}_{f}&=\mu_{f} -\frac{2}{3}\alpha G\sum_{f^{\prime}=u, d, s}\rho_{f^{\prime}}\ ,\label{eq:3f Effmass}
\end{aligned}
\end{equation}  
where we define $G' = (1-\alpha)G$, and $i, j, k $ represent even permutations of \( u, d, s \). From Eq.~(\ref{eq:3f Effmass}), it is evident that the introduction of the Fierz-transformed interactions contributes to the effective chemical potential and modifies the gap equations. 

From the grand canonical potential density of the interacting strange quark matter system
\begin{equation}
\begin{aligned}
&\Omega_M(T, \tilde{\mu})=\\
& -6 \sum_{f=u, d, s} \int \frac{\mathrm{d}^3 p}{(2 \pi)^3} \left[E_f +T \ln \left(1+e^{-\left(E_f+\tilde{\mu}_f\right) / T}\right)\right. \\
 &\left.+T \ln \left(1+e^{-\left(E_f-\tilde{\mu}_f\right) / T}\right)\right] 
 +2 G'\left(\sigma_u{ }^2+\sigma_d{ }^2+\sigma_s{ }^2\right) \\
 &-\frac{\alpha G}{3}\left(\rho_u+\rho_d+\rho_s\right)^2
-4 K \sigma_u \sigma_d \sigma_s+\Omega_0 \ ,  \label{eq: 3f grand thermo potential}
\end{aligned}
\end{equation}
the pressure can be calculated as follows:
\begin{equation}
\begin{aligned}
&P= -\Omega \\
& =6 \sum_{f=u, d, s} \int \frac{\mathrm{d}^3 p}{(2 \pi)^3} \left[E_f +T \ln \left(1+e^{-\left(E_f+\tilde{\mu}_f\right) / T}\right)\right. \\
 &\left.+T \ln \left(1+e^{-\left(E_f-\tilde{\mu}_f\right) / T}\right)\right] 
 -2G'\left(\sigma_u{ }^2+\sigma_d{ }^2+\sigma_s{ }^2\right) \\
 &+\frac{\alpha G}{3}\left(\rho_u+\rho_d+\rho_s\right)^2
+4 K \sigma_u \sigma_d \sigma_s- \Omega_0 \ ,\label{eq:3f pressure}
\end{aligned}
\end{equation}
and the energy density is
\begin{equation}
\begin{aligned} 
& \epsilon =-P+T S+\sum_f \mu_f n_f \\
&=6 \sum_{f=u, d, s} \int \frac{\mathrm{d}^3 p}{(2 \pi)^3} 
 \left[E_f\left(\bar{n}_f+n_f-1\right) \right]\\
&+2G'\left(\sigma_u{ }^2+\sigma_d{ }^2+\sigma_s{ }^2\right)+\frac{\alpha G}{3}\left(\rho_u+\rho_d+\rho_s\right)^2\\
&-4 K \sigma_u \sigma_d \sigma_s+\Omega_0\ . \label{eq:3f energy density}
\end{aligned}
\end{equation}
Here, we choose the constant $\Omega_0$ in Eqs.~(\ref{eq: 2f grand thermo potential}) and (\ref{eq: 3f grand thermo potential})  such that $P$ and $\epsilon$ vanish in the vacuum, i.e., $\Omega_M(0,0)=0$.

So far, we have successfully derived the thermodynamic
potential densities at finite chemical potential and temperature. In order to ensure thermodynamic consistency, the condensates must be obtained through
appropriate differentiation of the thermodynamic potential. The self-consistent solutions are those that correspond to the stationary points of the potential, which are defined by:
\begin{equation}
\frac{\delta \Omega}{\delta \sigma}=\frac{\delta \Omega}{\delta \tilde{\mu}}=0 \ , \label{eq:thermal satisfy}
\end{equation}
 from which we obtain the quark condensate expressed as
\begin{equation}
\begin{aligned}
\sigma_f=-\frac{N_cM_f}{\pi^2}  \int_{0}^{\Lambda} \frac{p^2}{E_f} \mathrm{d} p \left[1-n_f(T, \tilde{\mu}_f)-\bar{n}_f(T, \tilde{\mu}_f)\right] \ ,
\end{aligned}
\end{equation}
and the quark number densities are
\begin{equation}
\begin{aligned}
\rho_f=\frac{N_c}{\pi^2} \int p^2 \mathrm{d} p \left[n_f(T, \tilde{\mu})-\bar{n}_f(T, \tilde{\mu})\right]
\end{aligned}
\end{equation}
for $f=u, d, s$.

\section{Quark matter in $\beta$-equilibrium}\label{Sec3: quark matter at finite T}
A newborn neutron star forms in the aftermath of a successful supernova explosion, when the stellar remnant becomes gravitationally decoupled from the expanding ejecta and core temperatures reaches as high as $\sim50\mev$~\cite{2004Sci...304..536L,2005PrPNP..54..193W}.
In this work, we consider two distinct scenarios: an early stage in which neutrinos are trapped within the stellar interior, and a later stage after deleptonization, where neutrinos have escaped and their chemical potential vanishes. Although an isentropic description is generally considered more realistic for modeling protoquark stars, we adopt an isothermal approach to enable direct comparisons across different temperature cases.

In $\beta$-equilibrated matter, the contribution of leptons is included as a free Fermi gas in the calculation of the energy density, pressure, and entropy density. The relations among the chemical potentials of the various particle species without neutrinos under $\beta$-equilibrium are given by
\begin{equation}
\mu_s = \mu_d = \mu_u + \mu_e  \ . \label{beta equilibrium}
\end{equation} 
To describe stellar matter, the charge-neutrality of the quark matter should also be required,
\begin{equation}
\frac{2}{3}\rho_u-\frac{1}{3}\rho_d-\frac{1}{3}\rho_s-\rho_e=0 \:,  \label{charge-neutrality}
\end{equation}
and the baryon number conservation,
\begin{equation}
\frac{1}{3}\left(\rho_u + \rho_d + \rho_s\right) =\rho_{\rm B} \:, \label{eq30}
\end{equation}
is satisfied with $\rho_{\rm B}$ being the baryon number density. 
The baryon chemical potential satisfies $\mu_{\rm B} = \mu_u + \mu_d + \mu_s$ for strange quark matter, and $\mu_{\rm B} = \mu_u + 2\mu_d$ for nonstrange quark matter.

In earlier stages, when the neutrinos are still trapped in the interior of the star, Eq.~(\ref{beta equilibrium}) is replaced by 
\begin{equation}
\mu_s = \mu_d = \mu_u + \mu_e -\mu_{\nu e} \ , \label{beta equilibrium trapped}
\end{equation} 
and the lepton contribution $Y_l$ is set to be constant
\begin{equation}
(\rho_e+\rho_{\nu e})/\rho_B=Y_l  \ . \label{lepton fraction}
\end{equation} 
The lepton has no interactions, so the grand canonical potential density is the free grand canonical potential density. The lepton number densities are given by 
\begin{equation}
\begin{aligned}
\rho(\mu_e,T)=& 2 \int \frac{p^2 \mathrm{d} p}{2 \pi^2}\left[n_e(T, \mu)-\bar{n}_e(T, \mu)\right]\ ,\\
\rho(\mu_{\nu e},T)= & \int \frac{p^2 \mathrm{d} p}{2 \pi^2}\left[n_e(T, \mu)-\bar{n}_e(T, \mu)\right]\ .
\end{aligned}
\end{equation}

The EOS of dense warm matter is obtained by including both the contribution of quarks and leptons, namely the total energy density, pressure and entropy density are
\begin{equation}
\begin{aligned}
\epsilon=\epsilon_{\text{quark}}+\epsilon_{\text {lep }}, \quad P=P_{\text{quark}}+P_{\text {lep }}\ .
\end{aligned}
\end{equation}
The free energy density of the system is 
\begin{equation}
  \begin{aligned}
      f=\epsilon-TS~.\label{eq: freeE_relation}
\end{aligned}  
\end{equation}
In the case of isothermal scenario considered in the present work, where the temperature remains constant, the pressure can be expressed as 
\begin{equation}
  \begin{aligned}
      P=
      \rho_B^2 \frac{\partial}{\partial \rho_B}\left(\frac{f}{\rho_B}\right)_T~,\label{eq: freeeE}
\end{aligned}  
\end{equation}
which implies that the minimum of $f/\rho_B$ should occur at zero pressure~\cite{1967RPPh...30..615F,2013EPJC...73.2569D}. This condition, stemming from thermodynamic self-consistency, will be used as a criterion to further assess the EOS and to test its thermodynamic consistency in Sec.~\ref{Sec: star}.

\section{quark matter properties at finite temperature}\label{Sec: QM finite tem}
\subsection{fitting parameters}
Before doing the calculation, we should fit the NJL model parameters. Since the NJL model is not renormalizable, here, we use the three-momentum cutoff to regulate the ultraviolet divergences. After defining the new coupling constant $G^{\prime}$, and keeping the expression of gap equation the same as the widely used one in Eqs.~(\ref{eq: 2f gap Eq}) and~(\ref{eq:3f Effmass}), at zero temperature and chemical potential, apart from $\alpha$, the fixing of the model parameters is the same with the original version of the NJL model~\cite{1996PhRvC..53..410R}. For two-flavor NJL model, the $\Lambda_{\rm UV}$ and $G'$ are determined by fitting the experimental observables of pion decay constant and pion mass~\cite{2019ChPhC..43h4102W}. For the three-flavor case, we adopt the parameter sets of Rehberg, Klevansky, and H\"ufner (RKH)~\cite{1996PhRvC..53..410R}. This group sets $m_{u,d}=5.5\mev$, and the remaining four parameters, $m_s, \Lambda_{\mathrm{UV}}, G^{\prime}$, $K$, are chosen to reproduce the experimental data of the pion decay constant $f_{\pi}=92.4\mev$, the pion mass $m_{\pi}=135.0\mev$, the kaon mass $m_{K}=497.7\mev$, and the $\eta'$ meson mass $m_{\eta'}=957.8\mev$. 

\begin{figure*}
\centering
\includegraphics[width=0.48\textwidth]{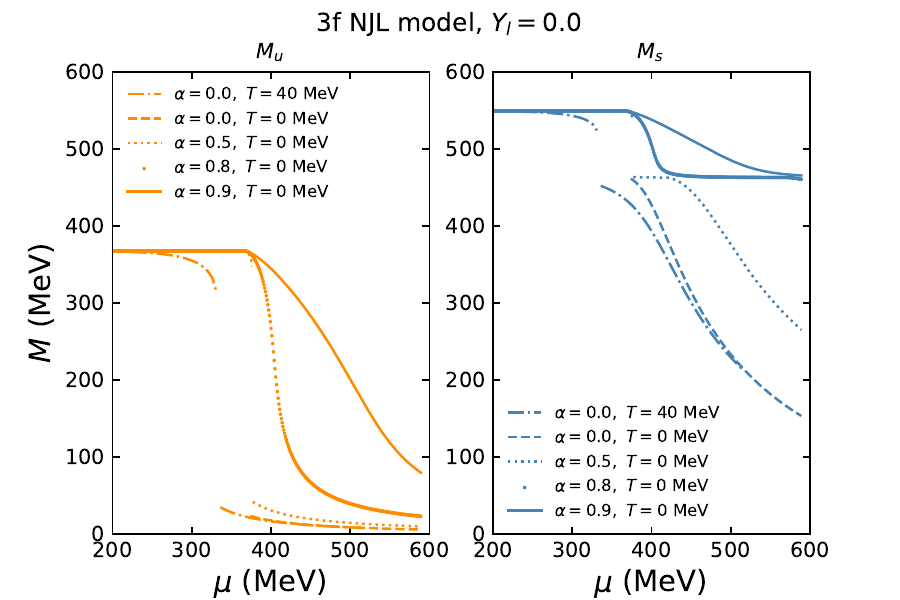}
\includegraphics[width=0.45\textwidth]{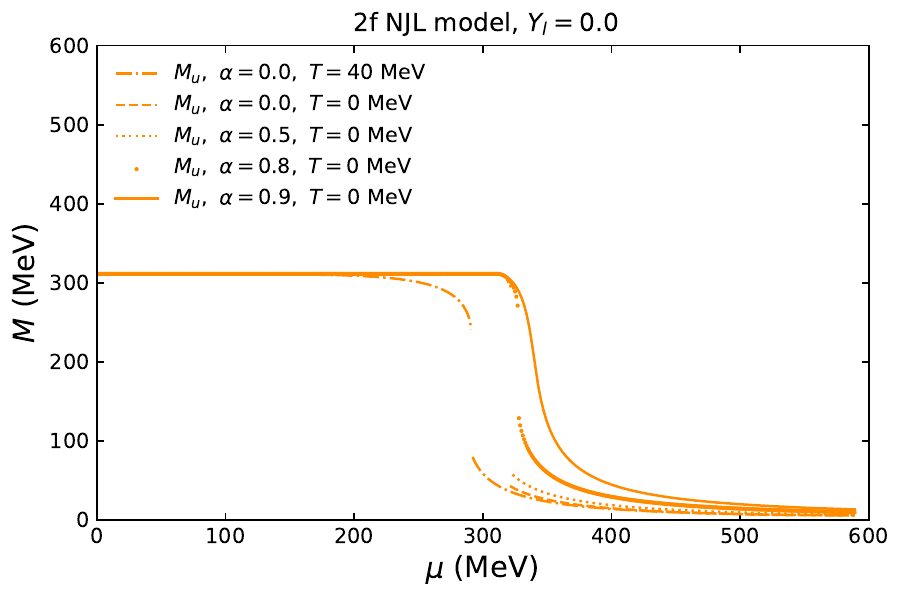}
\includegraphics[width=0.48\textwidth]{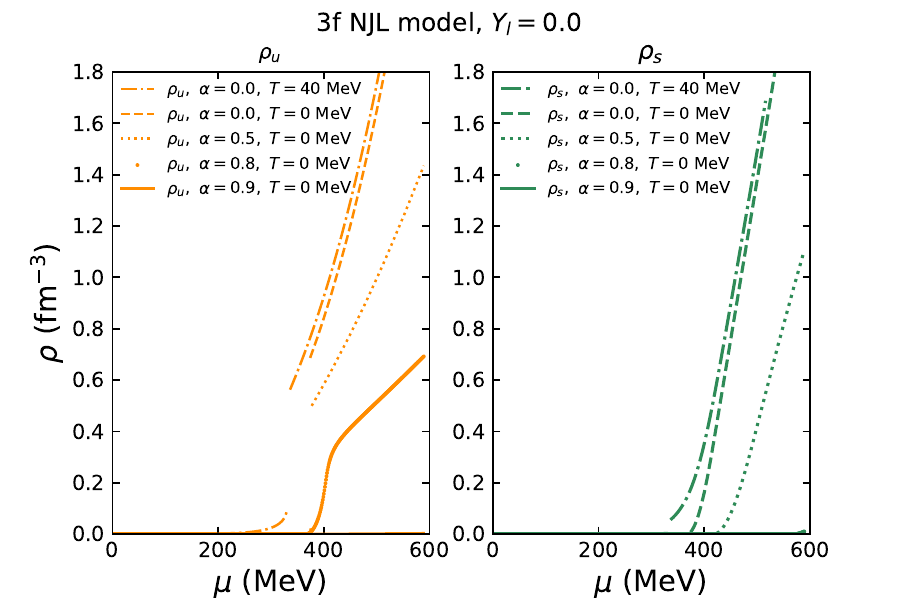}
\includegraphics[width=0.45\textwidth]{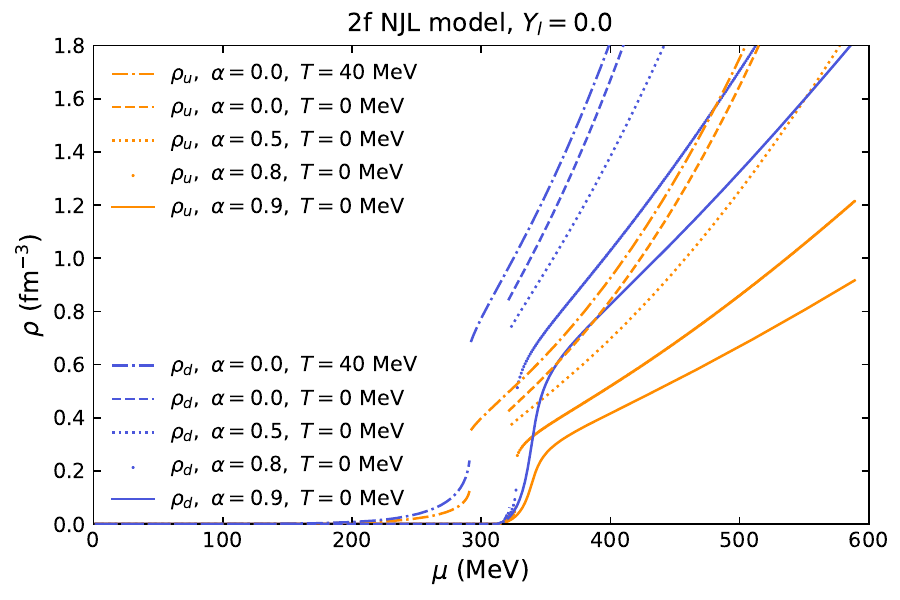}
\includegraphics[width=0.45\textwidth]{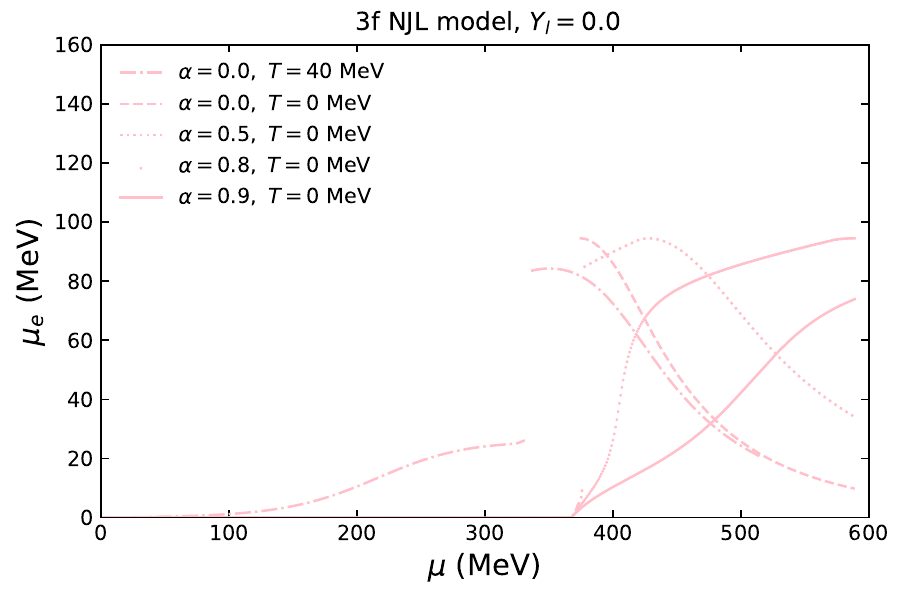} 
\includegraphics[width=0.45\textwidth]{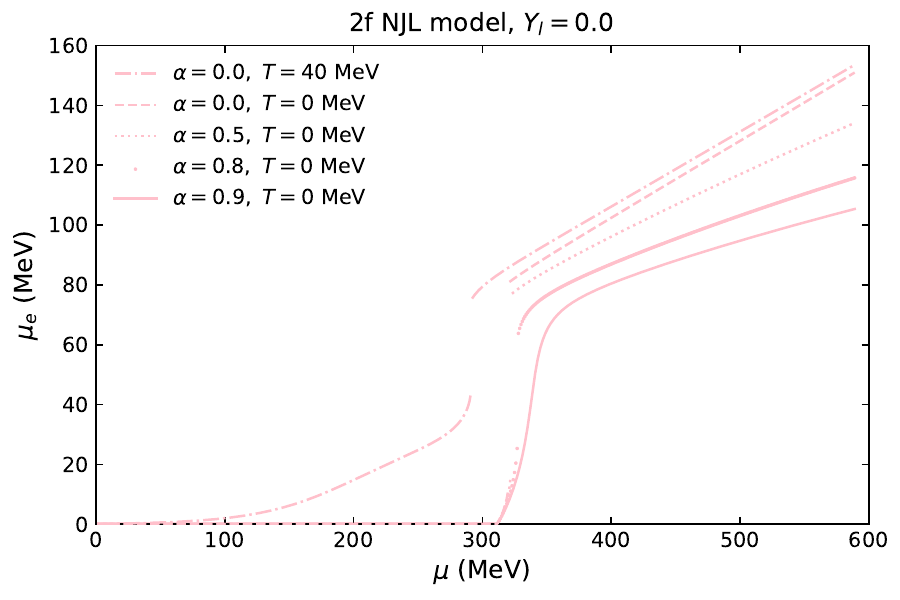}
\caption{Upper panel: The constituent quark masses of $u,\ d, \ s$ quarks versus quark chemical potential $\mu$ both for strange quark matter and nonstrange quark matter in the absence of neutrinos ($Y_l = 0.0$), under various values of $\alpha$ and temperature $T$. The results for several representative parameter sets are shown: ($\alpha=0.0$,\ $T=40$~MeV), ($\alpha=0.0$,\ $T=0$~MeV), ($\alpha=0.5$,\ $T=0$~MeV), ($\alpha=0.8$,\ $T=0$~MeV), ($\alpha=0.9$,\  $T=0$~MeV).  Middle panel: The corresponding results for quark number densities of $u,\ d, \ s$ quarks as functions of $\mu$. Lower panel: The electron chemical potential $\mu_e$ as a function of $\mu$.
}\label{fig:without neutrinos}
\end{figure*}
\begin{figure*}
\centering 
\includegraphics[width=0.48\textwidth]{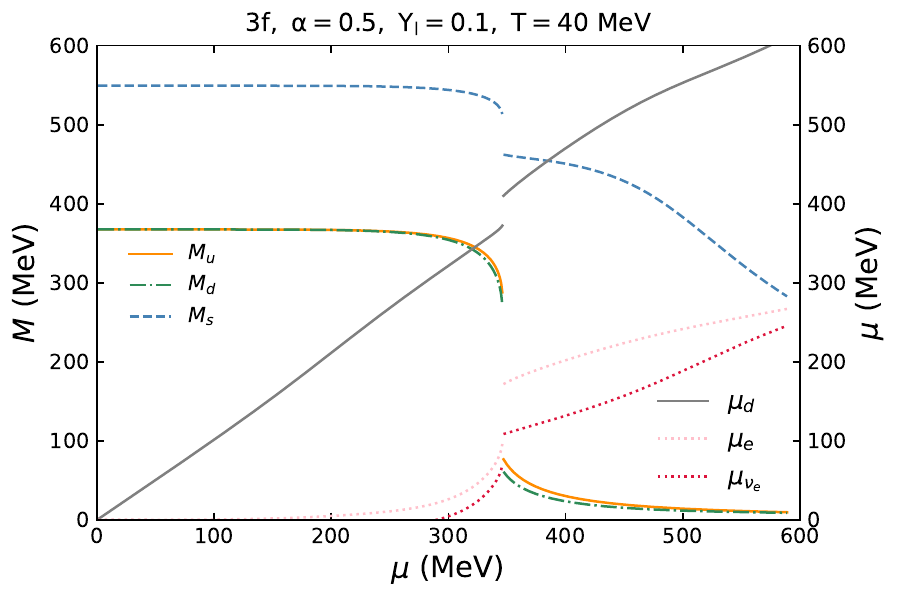}
\includegraphics[width=0.48\textwidth]{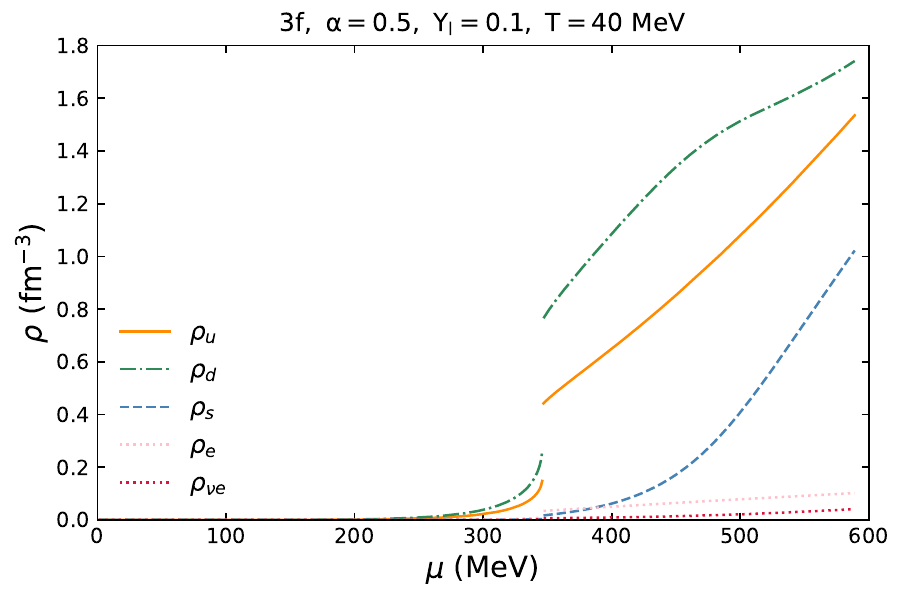}
\includegraphics[width=0.48\textwidth]{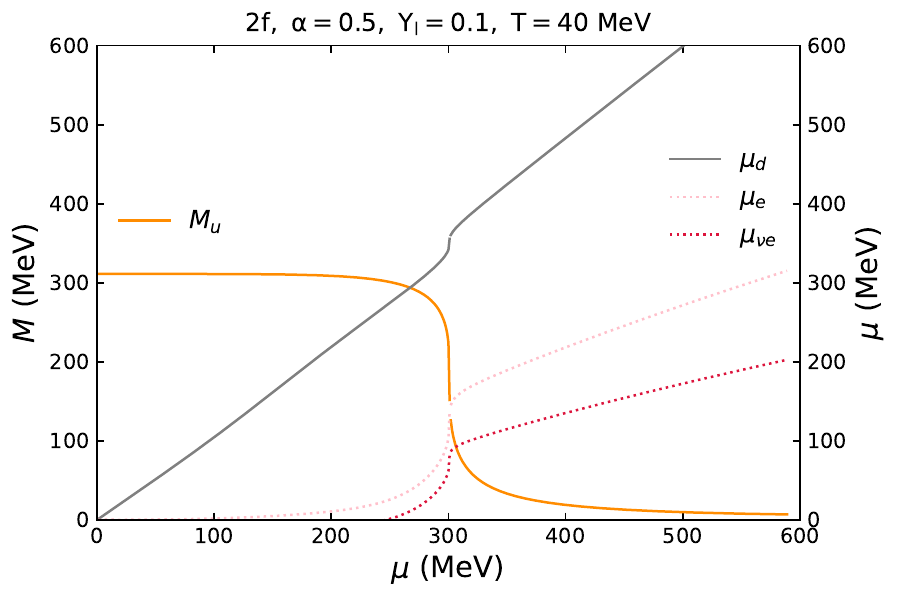}
\includegraphics[width=0.48\textwidth]{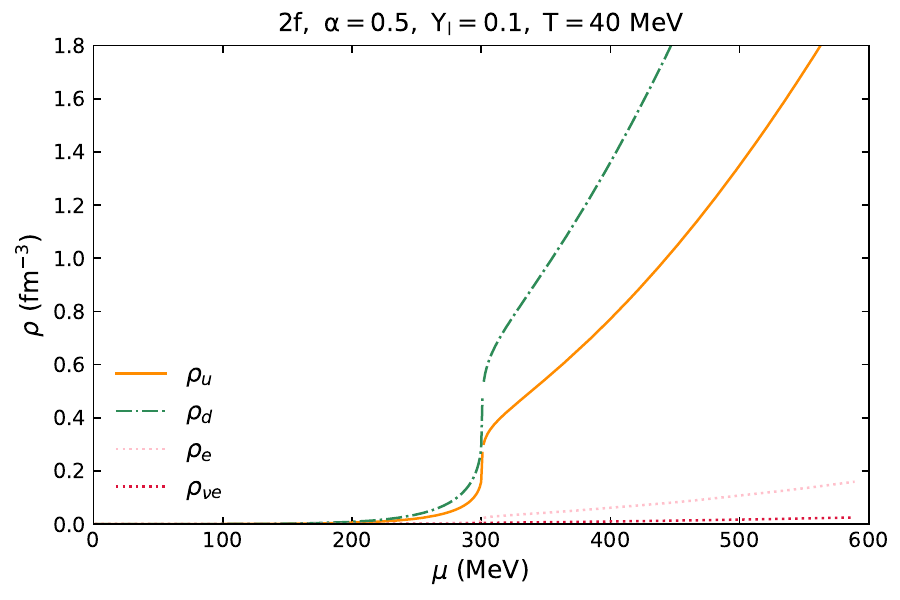}
\caption{The constituent quark masses $M_u$, $M_d$, and $M_s$, the electron chemical potential $\mu_e$, and the $d$ quark chemical potential $\mu_d$ are shown as functions of the $u$ quark chemical potential in the neutrino-trapped scenario, for a representative case with $\alpha = 0.5$, $Y_l = 0.1$, and $T = 40$~MeV. Results are presented for both strange and nonstrange quark matter.}\label{fig:lepton_typical case}
\end{figure*}

\begin{table}
\centering
\caption{NJL model parameters are shown. The units of the coupling constants $G'$ and $K$ are $\mathrm{MeV}^{-2}$ and $\mathrm{MeV}^{-5}$, respectively, and the other parameters have the units of $\mathrm{MeV}$.} 
         \vskip+2mm
\renewcommand\arraystretch{1.5} 
\begin{ruledtabular}
\begin{tabular*}{\hsize}{@{}@{\extracolsep{\fill}}lccccc@{}}
 & $m_{u}$ & $m_{s}$ & $\Lambda_{\mathrm{UV}}$ & $G'$ & $K$   \\
\hline Two-flavor~\cite{2019ChPhC..43h4102W}  & $5.0$ & /\multirow{2}{*} & $653.0$ & $4.930 \times 10^{-6}$ &
/ \\
\hline Three-flavor~\cite{1996PhRvC..53..410R} & $5.5$ & 140.7 & 602.3 & $5.058 \times 10^{-6}$ & $1.559 \times 10^{-13}$ \\
\end{tabular*}
\end{ruledtabular}
    \vspace{-0.4cm}
\label{table:1}
\end{table}

To obtain the EOS, the following set of equations must be solved self-consistently. For neutrino-free matter, the relevant system consists of the gap equations of the NJL model in Eq.~(\ref{eq: 2f gap Eq}) or Eq.~(\ref{eq:3f Effmass}), along with the conditions for $\beta$-equilibrium in Eq.~(\ref{beta equilibrium}) and charge neutrality in Eq.~(\ref{charge-neutrality}). These constraints reduce the number of independent chemical potentials to one. We choose the up quark chemical potential $\mu_u$ as the input quantity, with the remaining chemical potentials, $\mu_d$, $\mu_s$ and $\mu_e$, determined as functions of $\mu_u$. In the case of neutrino-trapped matter, the system includes the same NJL gap equations, together with the modified $\beta$ equilibrium condition in Eq.~(\ref{beta equilibrium trapped}), charge neutrality in Eq.~(\ref{charge-neutrality}), and a fixed lepton fraction in Eq.~(\ref{lepton fraction}). Again, only one independent chemical potential remains.

\subsection{without neutrinos}
In this section, we investigate the effects of exchange interactions and temperature on the properties of quark matter under conditions of $\beta$-equilibrium and charge neutrality, neglecting neutrino contributions. Figure~\ref{fig:without neutrinos} presents the constituent quark masses and quark number densities of $u,\ d, \ s$ quarks, as well as electron chemical potential calculated within the three-flavor and two-flavor NJL models. At zero temperature, when $\tilde{\mu}_f<M_f$, the quark condensate is independent of the quark chemical potential, and the constituent quark mass remains at its vacuum value. When  $\tilde{\mu}_f>M_f$, the chiral symmetry begins to restore. As the parameter $\alpha$ becomes nonzero, vector interactions are introduced self-consistently into the system. Due to the repulsive nature of these interactions in the exchange channel, the constituent quark mass decreases more gradually with increasing chemical potential $\mu$, particularly for the $s$ quark. This effect is clearly illustrated by comparing the cases of $\alpha=0.0, \ T=0.00$~MeV and $\alpha=0.5,\ T=0.00$~MeV. A further increase of $\alpha$ to 0.8 (0.9) enhances the role of vector interactions in the exchange channels, further suppresses chiral symmetry restoration, and drives the first-order chiral phase transition toward a crossover. In comparison with the NJL model without vector interactions ($\alpha=0.0, \ T=0.00$~MeV), increasing the temperature makes the quark condensate begins to restore at a relatively lower chemical potential.

As expected, at zero temperature, the quark number density stays zero when $\mu$ is smaller than the constituent quark mass $M_{u,d}$ and $M_s$. Once $\mu$ exceeds the corresponding constituent quark mass, the quark number density exhibits a discontinuous jump to a nonzero value, signaling a first-order phase transition. In the two-flavor quark matter system, the restoration of the quark condensate leads to an increase in the up and down quark number densities $\rho_{u,d}$. Due to the charge neutrality condition, this necessitates an increase in the electron fraction, and thus, the electron chemical potential $\mu_e$ rises with increasing $\mu$. 
In contrast, the behavior of $\mu_e$ in strange quark matter is more intricate. Increase the value of $\alpha$ enhances the vector interaction strength, and these repulsive interactions suppress the restoration of the strange quark condensate, thereby reducing the strange quark number density $\rho_s$. Consequently, particularly for $\mu \gtrsim 400$~MeV, a larger value of $\alpha$ leads to a higher electron chemical potential, compensating for the reduced negative charge from strange quarks in order to maintain overall charge neutrality. In particular, for $\alpha=0.8$, the strange quark condensate is rarely restored, yielding a system with a negligible strange quark population and, consequently, a significantly elevated  $\mu_e$ at high $\mu$.

\begin{figure*}
\centering
\includegraphics[width=0.47\textwidth]{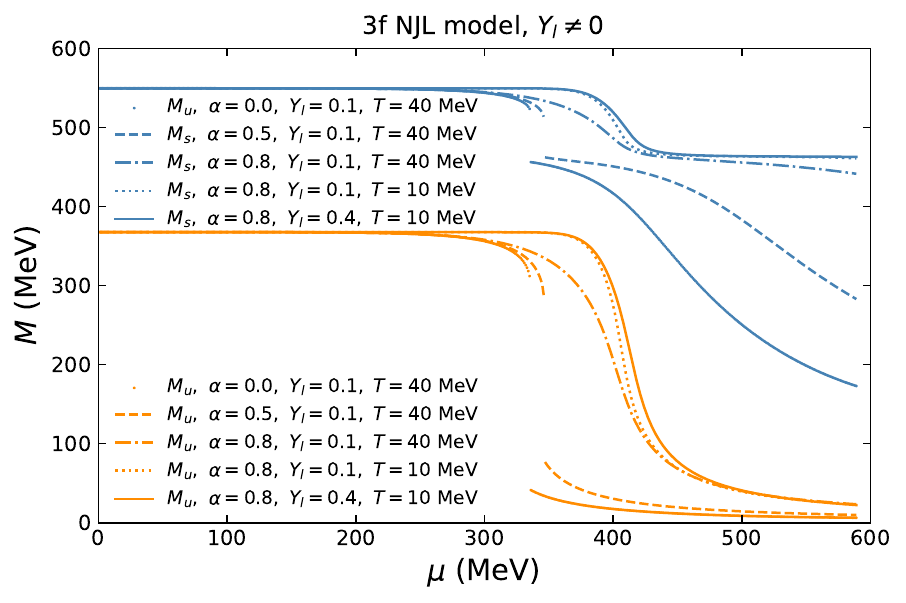}
\includegraphics[width=0.47\textwidth]{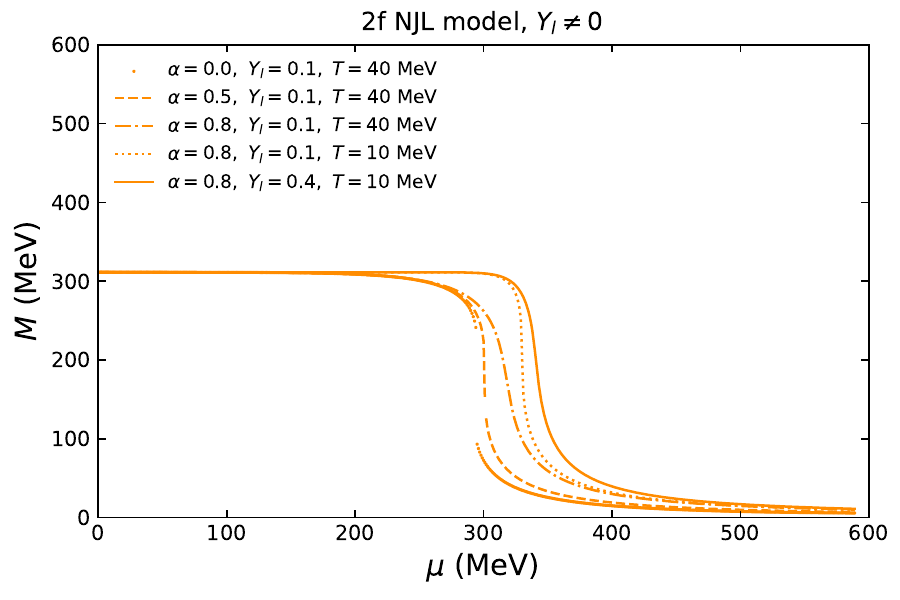}
\includegraphics[width=0.47\textwidth]{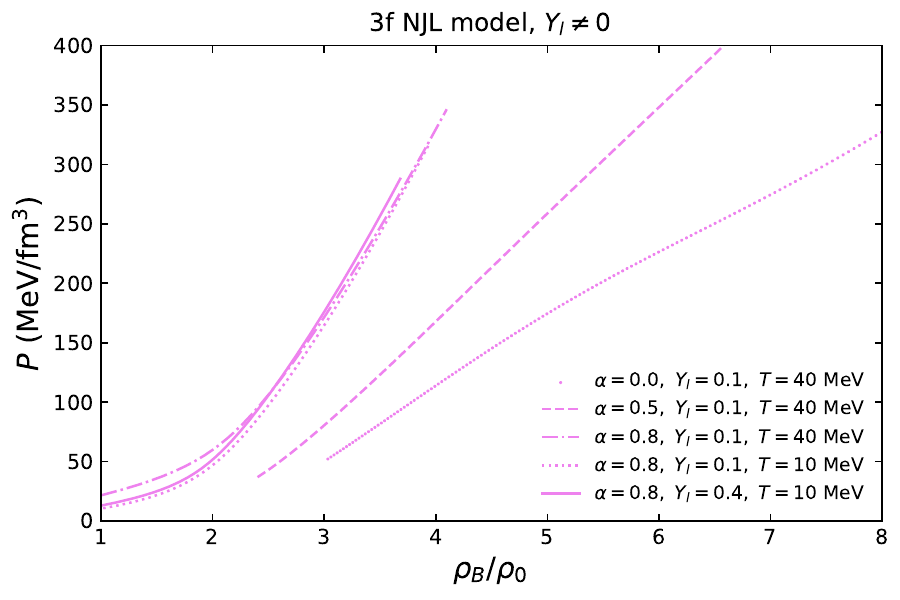} 
\includegraphics[width=0.47\textwidth]{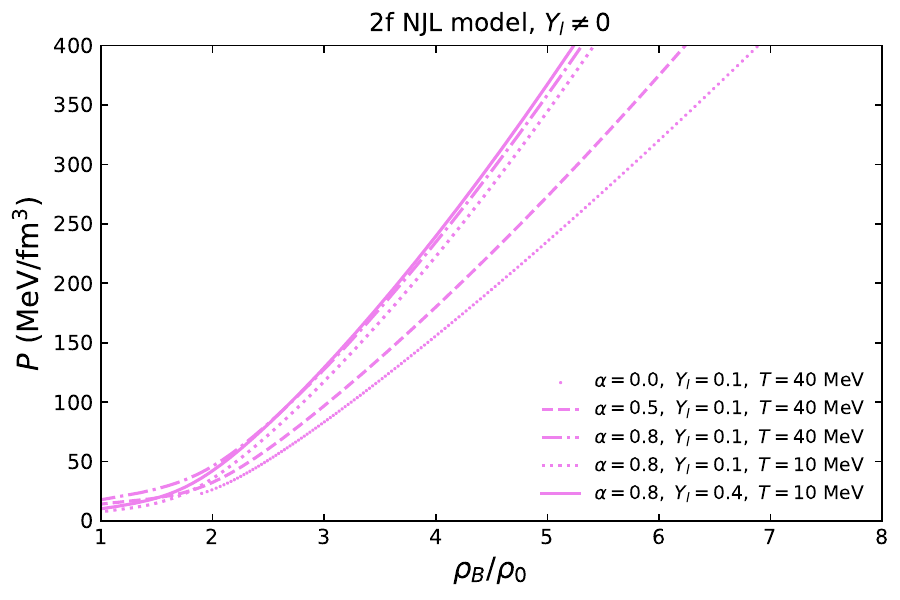} 
\caption{The constituent quark masses of $u,\ d,\ s$ quarks versus quark chemical potential $\mu$ for various choices of the three- and two-flavor NJL model parameters with neutrino trapped scenario for the study of their effects (see text for details). Lower panel: The pressure as a function of baryon number density, expressed in units of the nuclear saturation density $\rho_0$, for both strange quark matter and nonstrange quark matter. 
}\label{fig:compare_2f_3f}
\end{figure*}

\subsection{neutrino trapped}
To investigate the properties of protoquark stars, we assume that neutrinos are trapped in the matter. In the following analysis, we consider a temperature range of $T = 0-40\mev$ and a total lepton fraction of $Y_l = 0.1-0.4$, consistent with conditions expected in compact stars during the core-collapse phase of massive stellar evolution~\cite{1986ApJ...307..178B}. We will demonstrate that the strength of exchange interactions and the vacuum pressure play a more significant role than the lepton fraction and temperature in determining the stiffness of the EOS. The underlying reasons for this behavior will be discussed later.

For neutrino-trapped matter, we solve Eqs.~(\ref{eq: 2f gap Eq} or~\ref{eq:3f Effmass}), (\ref{charge-neutrality}), (\ref{beta equilibrium trapped}), and (\ref{lepton fraction}) self-consistently, taking the chemical potential of the $u$ quark as the input quantity. The resulting physical solutions for two- and three-flavor quark matter, corresponding to a representative case with $\alpha = 0.5$, $Y_l = 0.1$, and $T = 40\mev$, are shown in Fig.~\ref{fig:lepton_typical case}. For three-flavor quark matter, When $\mu > 220~\mathrm{MeV}$, the $u$ quark condensate begins to melt slightly, accompanied by the appearance of a small population of $u$ and $d$ quarks. Owing to flavor-mixing effects and the associated reduction of $M_u$ and $M_d$, the constituent mass of the $s$ quark also decreases slightly. At $\mu = 351~\mathrm{MeV}$, a first-order phase transition occurs, marked by a discontinuous drop in $M_u$, $M_d$, and $M_s$. The $M_u$ differs slightly from the $d$ quark mass due to the $\beta$-equilibrium. For $\mu > 365~\mathrm{MeV}$, the $ud$ quark masses are fully restored to their current masses, and the flavor-mixing effect becomes negligible. As the quark chemical potential increases further, the strange quark condensate continues to decrease, and the number density of $s$ quarks increases. In contrast, the two-flavor case exhibits a relatively low constituent quark mass of approximately $311~\mathrm{MeV}$, generated via spontaneous chiral symmetry breaking, with only a weak discontinuity in the $M(\mu)$ relation. Due to the absence of strange quarks, both the electron chemical potential $\mu_e$ and the neutrino chemical potential $\mu_{\nu _e}$ increase monotonically with $\mu_u$, driven by the requirements of charge neutrality and $\beta$-equilibrium.

In Fig.~\ref{fig:compare_2f_3f}, we examine the effects of lepton fraction $Y_l$, temperature $T$, and the strength of exchange interactions--characterized by the parameter $\alpha$--on both strange and nonstrange quark matter. We present the constituent quark masses $M_u$ and $M_s$ as functions of the quark chemical potential $\mu$, along with the pressure as a function of baryon number density $\rho_B$ for various parameter sets. In the lower panel of Fig.~\ref{fig:compare_2f_3f}, for the first-order chiral phase transition, we consider only the chirally restored quark matter phase predicted by the NJL model to be reliable; therefore, the results are shown starting from $\rho_0$. 
Under the same conditions of $Y_l = 0.1$ and $ T= 40\mev$, comparing stronger vector interactions with $\alpha = 0.5$ to those with $\alpha = 0.0$ shows that the former shifts the onset of the first-order phase transition to higher quark chemical potentials and significantly suppresses the restoration of constituent quark masses at high chemical potentials, especially that of the strange quark. A similar trend is observed in the case of nonstrange quark matter. As shown in the lower panel, the EOS becomes stiffer for larger $\alpha$, reflecting a stronger contribution from vector interactions in the exchange channels. 
Compared to the case of $\alpha = 0.8$, $Y_l = 0.1$, and $T = 10\mev$, increasing the temperature to $T = 40\mev$ accelerates the melting of the constituent quark masses before $\mu\sim 400\mev$, causing the strange quark phase to emerge at lower chemical potentials. This behavior can also be found for nonstrange quark matter. The temperature has only a slight effect on the stiffness of the EOS, as can be seen from the $P(\rho_B)$ relations. 
For fixed $\alpha = 0.8$ and $T = 40\mev$, increasing the lepton fraction from $Y_l = 0.1$ to $Y_l = 0.4$ slightly suppresses the restoration of constituent masses and induces a modest change in the stiffness the EOS of quark matter. A higher lepton fraction implies a larger electron population, which, due to the charge neutrality condition, suppresses the abundance of other negatively charged particles, such as the $s$ quark. As a result, a higher $Y_l$ leads to a reduced fraction of $s$ quarks, thereby hindering the restoration of their constituent mass. A analogous effect was also reported in Ref.~\cite{2009PhRvD..79d3012Y}, which investigated the effects of leptons on proto-neutron stars.

\begin{figure*}
\centering
\includegraphics[width=0.49\textwidth]{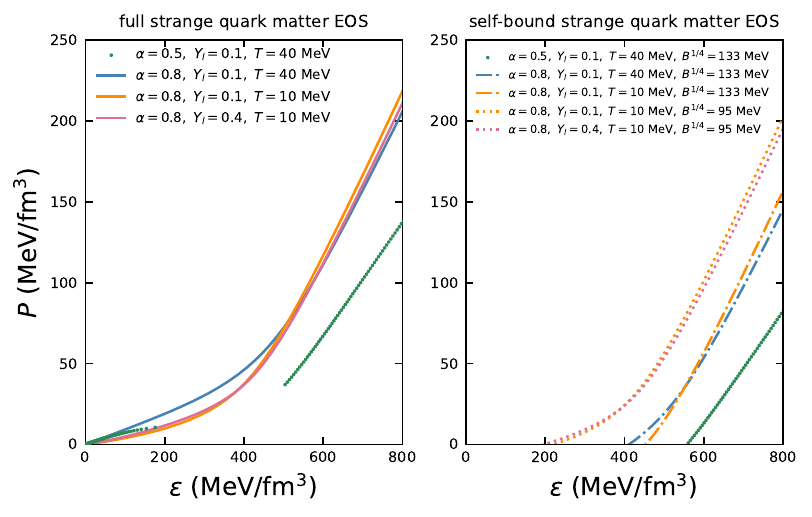}
\includegraphics[width=0.49\textwidth]{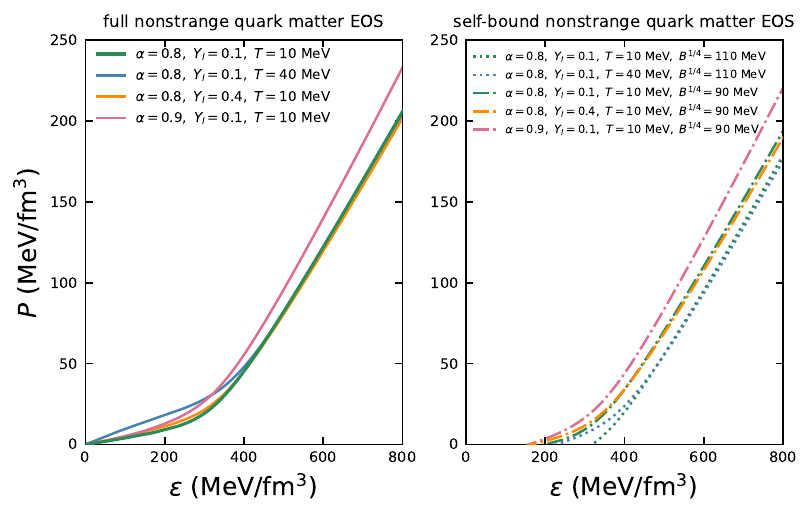}
\includegraphics[width=0.49\textwidth]{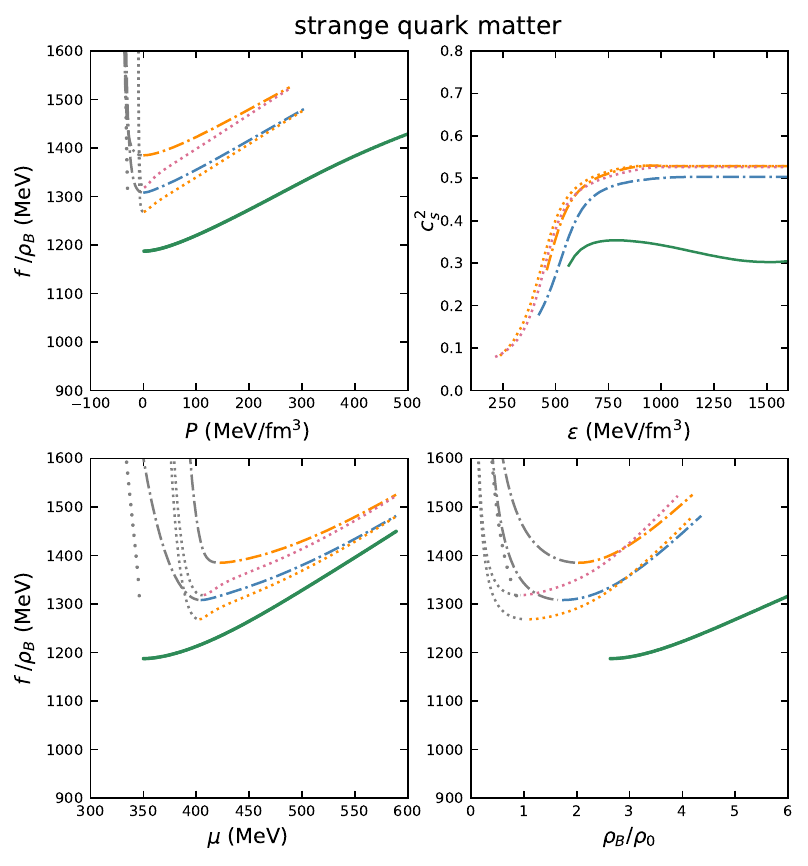}
\includegraphics[width=0.49\textwidth]{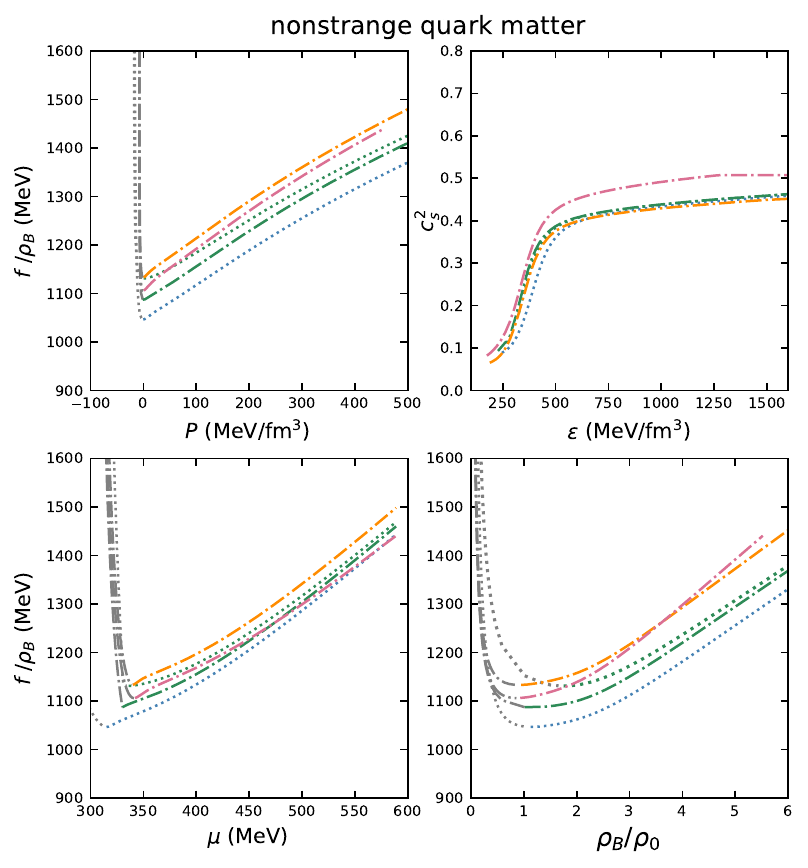}
\caption{Upper panel: EOS for several representative parameter sets calculated within NJL model. Results are shown for the three-flavor case with ($\alpha=0.5$, $Y_l=0.1$,\ $T=40$~MeV, \ $B^{1/4}=133\mev$), ($\alpha=0.8$,\ $Y_l=0.1$,\ $T=40$~MeV, $B^{1/4}=133\mev$), ($\alpha=0.8$,\ $Y_l=0.1$, \ $T=10$~MeV,\ $B^{1/4}=133\mev$), ($\alpha=0.8$,\ $Y_l=0.1$, \ $T=10$~MeV,\ $B^{1/4}=95\mev$), ($\alpha=0.8$,\ $Y_l=0.4$, \ $T=10$~MeV,\ $B^{1/4}=95\mev$), as well as for the two-flavor case with ($\alpha=0.8$,\ $Y_l=0.1$, \ $T=10$~MeV,\ $B^{1/4}=110\mev$), ($\alpha=0.8$,\ $Y_l=0.1$, \ $T=40$~MeV,\ $B^{1/4}=110\mev$), ($\alpha=0.8$,\ $Y_l=0.1$, \ $T=10$~MeV,\ $B^{1/4}=90\mev$), ($\alpha=0.8$,\ $Y_l=0.4$, \ $T=10$~MeV,\ $B^{1/4}=90\mev$), ($\alpha=0.9$,\ $Y_l=0.1$, \ $T=10$~MeV,\ $B^{1/4}=90\mev$). The panel shows the full EOS, featuring the phase with unrestored chiral symmetry on the left and the restored quark phase corresponding to self-bound strange quark matter with a nonzero vacuum pressure $B$ on the right. Middle panel: Free energy per baryon $f/\rho_B$ as a function of pressure $P$, and the squared speed of sound $c_s^2$ as a function of energy density $\epsilon$, corresponding to the parameter sets shown in the upper panel for self-bound strange and nonstrange quark matter. The gray segments of the $f/\rho_B$ curves  represent thermodynamically unstable EOSs~\cite{1967RPPh...30..615F}.
Lower panel: The $f/\rho_B$ as functions of $\mu$ and $\rho_B/\rho_0$ for strange and nonstrange quark matter.
}\label{fig:3f_EOS_cs} 
\end{figure*}

\section{Stellar matter and protoquark star properties}\label{Sec: star}
\begin{figure*}
\centering
\includegraphics[width=0.90\textwidth]{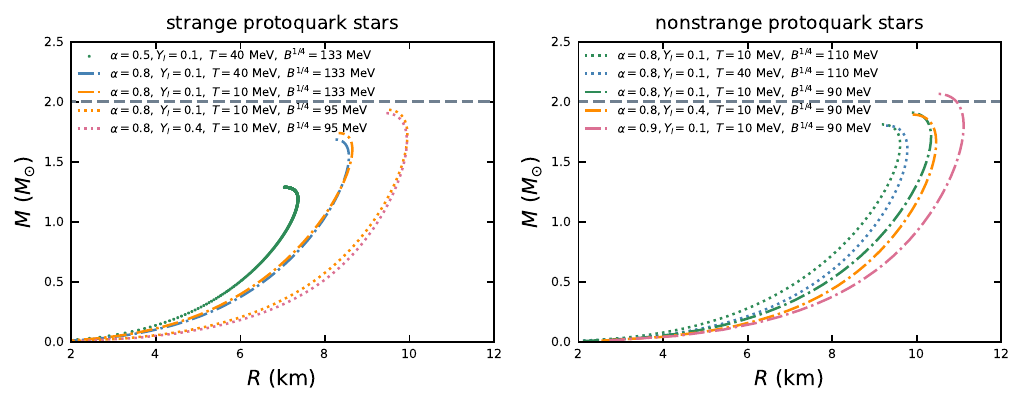}
\caption{The mass-radius (M-R) relations for strange and nonstrange protoquark stars corresponding to the same parameter sets shown in Fig.~\ref{fig:3f_EOS_cs}.
}\label{fig:3f_MR} 
\end{figure*}

According to the Bodmer-Witten hypothesis~\cite{1971PhRvD...4.1601B,1984PhRvD..30..272W}, absolutely stable quark matter must have an energy per baryon lower than that of the $^{56}\mathrm{Fe}$ nucleus at zero pressure and zero temperature. Accordingly, for self-bound strange quark matter, the condition $(E/A)_{\mathrm{uds}} \leq 930~\mathrm{MeV}$ must be satisfied for the hypothesis of absolutely stable strange matter to hold. In contrast, bulk nonstrange quark matter should have an energy per baryon higher than that of the confined phase, i.e., $(E/A)_{\mathrm{ud}} \geq 930~\mathrm{MeV}$. For absolutely stable nonstrange quark matter, the stability condition is reversed: $(E/A)_{\mathrm{ud}} \leq 930~\mathrm{MeV} \leq (E/A)_{\mathrm{uds}}$. In our previous work using the modified NJL model with exchange interactions~\cite{2022PhRvD.105l3004Y}, we demonstrated that there is sufficient parameter space to support both absolutely stable nonstrange and strange quark matter at zero temperature. However, at finite temperature, these conditions are not necessarily required to be satisfied.

In the following, we study the influence of the exchange interaction parameter $\alpha$, lepton fraction $Y_l$, and temperature $T$ on the interacting quark matter EOS and the structure of protoquark stars. As shown in Sec.~\ref{Sec: QM finite tem}, increasing the strength of the exchange channel $\alpha$ alters the nature of the chiral phase transition from first-order to a crossover. Since the relationship between chiral symmetry restoration and deconfinement remains unclear, in this work, we assume that deconfinement occurs simultaneously with the chiral phase transition. Therefore, to effectively mimic confinement at low densities, a finite vacuum pressure is introduced. According to the properties of the chiral phase transition, we adopt different prescriptions for defining the vacuum pressure $-B$:
\begin{itemize}
\item For a first-order phase transition at low temperature and weak vector interaction, the chiral symmetry breaking phase cannot be reliably described by the NJL model. Therefore, we consider only the chiral restored quark phase after the transition and choose the minimum vacuum pressure $B_0$ such that the pressure vanishes at a finite energy density, as required for modeling self-bound quark stars. Any other choice of $B$ should be larger than $B_0$. For example, in the upper panel of Fig.~\ref{fig:3f_EOS_cs}, for the case of $\alpha = 0.5$, $Y_l = 0.1$, and $T = 40~\mathrm{MeV}$, the chiral phase transition is of first order, resulting in a discontinuity in the energy density of the EOS. To ensure that the quark star is composed of chirally restored quark matter, the $B^{1/4}$ must exceed $133~\mathrm{MeV}$, allowing the quark matter EOS to reach zero pressure at a finite energy density—thus ensuring the star is self-bound at the surface.

\item For a crossover chiral transition occurring at relatively high temperatures and with strong vector interactions in the exchange channel, the transition becomes smooth. In this case, although the choice of vacuum pressure is more flexible, selecting a sufficiently small vacuum pressure ensures that the EOS yields a finite energy density at zero pressure; nevertheless, a minimum value of $B$ is still required to guarantee that the surface baryon density $\rho_B$ exceeds the nuclear saturation density $\rho_0$. The results below will indicate that the vacuum pressure significantly influences the structure of protoquark stars, suggesting that tighter constraints in the future on this parameter are essential for a deeper understanding of the quark matter EOS and compact star configurations.
\end{itemize}

\subsection{Strange stellar quark matter and stars}\label{Sec: strange PQS}
In the left panel of Fig.~\ref{fig:3f_EOS_cs}, we present the EOS at finite temperature with neutrino trapping, and the speed of sound $c_s^2$ for stellar matter, along with the behavior of the free energy per baryon $f/\rho_B$ for protoquark stars. The full EOSs represent results derived from the NJL model, including the chirally broken phase. To model self-bound quark matter that can form quark stars, a vacuum pressure $B$ is added to ensure the system reaches zero pressure at a finite density. For the case of $\alpha = 0.5$, $Y_l = 0.1$, and $T = 10\,\mathrm{MeV}$, a discontinuous jump in energy density appears, characteristic of a first-order phase transition. In this scenario, the vacuum pressure must be at least $B_0^{1/4} = 133\,\mathrm{MeV}$ to ensure that the pressure vanishes at the surface of the compact star. However, when the chiral phase transition becomes a crossover, the required vacuum pressure $B$ can be smaller than $B_0$.

In general, increasing the value of $\alpha$ enhances the repulsive vector interactions, leading to a stiffer EOS significantly, as clearly illustrated by the comparison of $\alpha = 0.5$ and $\alpha = 0.8$ in the behavior of the sound velocity. Strong vector interactions are necessary to support massive compact stars~\cite{2005PhR...407..205B}. For large values of $\alpha$, such as $\alpha = 0.8$, the chiral phase transition becomes a crossover. This smooth transition typically results in a stiffer EOS compared to the case of a first-order phase transition. For the case of $\alpha=0.8, Y_l=0.1,\ B^{1/4}=133\mev$, increasing the temperature from $T=10\mev$ to $40\mev$ softens the EOS, which can be seen from the behavior of $c_s^2$, leading to a modest reduction in the maximum mass $M_\mathrm{TOV}$ in Fig.~\ref{fig:3f_MR}. This reduction arises because the thermal contribution to the energy density exceeds that to the pressure. Ref.~\cite{2024PhRvD.110l3021P} has also found that the maximum mass decreases with increasing temperature in the isothermal scenario within the quasiparticle model incorporating vector interactions.
When the total lepton fraction $Y_l$ increases from 0.1 to 0.4 (with $\alpha = 0.8$, $T = 10~\mathrm{MeV}$, and $B^{1/4} = 95~\mathrm{MeV}$), the EOS with a larger lepton fraction becomes slightly softer, particularly at higher energy densities. At $\alpha = 0.8$, $Y_l = 0.1$, and $T = 10\mev$, a representative vacuum pressure of $B^{1/4} = 95~\mev$ yields a maximum mass of $1.93\msun$ for the strange protoquark stars. Further increasing the vacuum pressure to $B^{1/4} = 133\mev$ results in a much softer EOS. The speed of sound directly characterizes the stiffness of the EOS. It exhibits a rapid increase at low densities, then saturates at $c_s^2\sim0.5$, when the chiral transition undergoes a crossover at higher densities. The green dots indicate the first-order phase transition scenario, which exhibits a soft EOS with $c_s^2\sim0.3$ across the entire density range.

As discussed in Sec.~\ref{Sec3: quark matter at finite T}, thermodynamic consistency requires that the minimum of the free energy per baryon, $f/\rho_B$, occur at zero pressure. The behavior of $f/\rho_B$ as a function of $P$ clearly demonstrates thermodynamic consistency within the present model and also provides insight into the nature of the phase transition. For $P>0$, the value of $f/\rho_B$ increases with increasing chemical potential $\mu$ (or $\rho_B/\rho_0$ and $P$). 
In the case of a first-order phase transition, typically realized for parameters $\alpha = 0.8$, $Y_l = 0.1$, $T = 40$~MeV, and $B^{1/4} = 133$~MeV, $f/\rho_B$ exhibits a discontinuity, whereas a smooth crossover leads to a continuous behavior. As $B^{1/4}$ decreases, the baryon number density $\rho_B/\rho_0$ corresponding to the minimum of $f/\rho_B$ also decreases. However, this density must be at least larger than the nuclear saturation density $\rho_0$ to remain consistent with terrestrial nuclear experiments, thereby providing an additional constraint on the vacuum pressure $B$. In particular, for strange quark matter with $\alpha = 0.8$, $Y_l = 0.4$, and $T = 10$~MeV, a minimum value of $B^{1/4}_{\rm min} = 94$~MeV is required to ensure that the baryon number density at the stellar surface satisfies $\rho_B > \rho_0$ when $P = 0$. To the best of our knowledge, this physically motivated constraint on the phenomenological vacuum pressure $B$ has not been explicitly emphasized in previous studies.

It should be noted that the mass constraints derived from NICER observations~\cite{2019ApJ...887L..24M,2019ApJ...887L..21R,2021ApJ...918L..28M,2021ApJ...918L..27R} can only be applied to cold compact stars, meanwhile the gravitational-wave constraints from GW170817 were obtained from the late inspiral phase of the merger under the assumption of zero temperature ~\cite{2017PhRvL.119p1101A,2018PhRvL.121p1101A}. Therefore, they are not directly applicable to hot compact stars. The precise temperature of neutron stars during the inspiral phase remains an open question~\cite{2022Univ....8..395L}. In future work, investigating the influence of finite-temperature effects on tidal deformability and the I-Love-Q relations within the framework of the modified NJL model will further improve our understanding of dense matter and the associated astrophysical phenomena.

\subsection{Nonstrange stellar quark matter and stars}\label{Sec: nonstrange PQS}
The right panel of Fig.~\ref{fig:3f_EOS_cs} shows the EOS of nonstrange quark matter derived from our modified NJL model, together with the sound velocity and the free energy per baryon for several representative parameter sets. For both $\alpha=0.8$ and $\alpha=0.9$, the chiral phase transition is smooth, yielding continuous EOSs. To ensure that finite-temperature nonstrange quark matter is self-bound, i.e. capable of forming a quark star, we add a vacuum pressure term $B$. The minima of $f/\rho_B$ all occur at zero pressure. To ensure that the baryon number density at the surface of the quark star exceeds $\rho_0$, the minimum vacuum pressure is $B^{1/4}{\rm min} = 80$~MeV for $\alpha = 0.8$, $Y_l = 0.1$, and $T = 10$~MeV. Increasing the strength of the exchange interaction to $\alpha = 0.9$ raises the minimum vacuum pressure to $B^{1/4}{\rm min} = 90$~MeV. The speed of sound quickly rises to approximately $c_s^2=0.4$ at an energy density of around $500$~MeV/fm$^3$, with only minor changes to $c_s^2=0.46$ and $c_s^2=0.51$ for values of $\alpha$ between 0.8 and 0.9, respectively, at higher energy densities.

At $\alpha=0.8,\ T=10\mev,\ B^{1/4}=90\mev$, raising the lepton fraction from $Y_l=0.1$ to $Y_l=0.4$ softens the EOS at large densities, resulting in a relatively smaller maximum mass, as shown in the right panel of Fig.~\ref{fig:3f_MR}. 
Increasing the exchange channel strength from $\alpha=0.8$ to $\alpha=0.9$ produces a marked stiffening of the EOS, leading to $M_{\rm TOV}=2.07$ for the maximum mass. Increasing the vacuum pressure from $B^{1/4}=90\mev$ to $B^{1/4}=110\mev$ results in a lower maximum star mass. Adding large thermal effects to the nonstrange quark matter system slightly softens the EOS, which induces a lower $M_{\rm TOV}$ but larger radius. Compared with strange quark matter, the temperature effect on the stellar radius is somewhat more pronounced in the nonstrange case.



\section{Summary}\label{Sec: summary}
In this work, we investigate protoquark stars by combining the EOS of quark matter at finite temperature with a modified NJL model that incorporates exchange interactions, weighted by a parameter $\alpha$, for both two- and three-flavor cases. Unlike previous studies that introduce vector interactions manually, our approach uses the Fierz transformation to include scalar and vector contributions from the exchange channels self-consistently. Within this framework, we find that increasing the exchange interaction strength $\alpha$ modifies the chiral phase transition from a first-order transition to a crossover. Treating the vacuum pressure as a phenomenological parameter, we find that a first-order chiral phase transition within the NJL model necessitates a relatively large vacuum pressure to ensure vanishing pressure at the stellar surface, thereby yielding a softer EOS. The free energy per baryon, as well as the value of its minimum, decreases with a smaller vacuum pressure. Combining this with the requirement that the baryon number density at the stellar surface must be at least larger than $n_0$ provides a lower bound on the vacuum pressure $B$. In the isothermal scenario, the equation of state is almost insensitive to variations in temperature and lepton fraction.

Looking ahead, a more realistic treatment of protoquark stars should incorporate isotropic conditions and magnetic field effects~\cite{2020JPhG...47h5201C,2022PhRvC.105d5806C}. Beyond the normal quark phase, the potential role of a color superconducting phase and pion condensation phase at finite temperature is also particularly intriguing, especially in scenarios where high baryon densities are realized in the early Universe due to a large lepton number~\cite{2021PhRvL.126a2701V,2025arXiv250706518F}. Within the same finite-temperature NJL framework, it would be highly interesting to investigate the influence of diquark condensates on the evolution of protoquark stars. The corresponding results would be crucial for understanding the cooling mechanism of the compact stars, and could shed light on the behavior of dense matter and its broader cosmological implications.

\medskip
\acknowledgments
We thank Prof. Chen Zhang, Prof. Pengcheng Chu, Dr. Bingjun Zuo, and the PKU Pulsar Group for their useful discussions. We also would like to thank the referee for the insightful and valuable comments and suggestions. This work is supported by the National Natural Science Foundation of China (Nos. 12003047, 12133003), the Strategic Priority Research Program of the Chinese Academy of Sciences (No. XDB0550300). Wen-Li Yuan is supported by the China Postdoctoral Science Foundation (Grant. No. 2025M773418). Nobutoshi Yasutake is supported by the JSPS KAKENHI (Grant No. 24K07054).

\end{document}